\documentclass[journal,10pt]{IEEEtran}

\usepackage{stmaryrd}
\usepackage{amsmath}
\usepackage{multirow}
\usepackage{enumerate}
\usepackage{amsfonts}
\usepackage{color}
\usepackage{amssymb}
\usepackage{threeparttable}
\usepackage{array}
\usepackage[squaren,Gray]{SIunits}
\usepackage{enumitem}
\usepackage{slashbox}

\ifCLASSINFOpdf
   \usepackage[pdftex]{graphicx}
  \DeclareGraphicsExtensions{.pdf,.jpeg,.png}
\else
   \usepackage[dvips]{graphicx}
  \DeclareGraphicsExtensions{.eps}
\fi

\ifCLASSOPTIONcompsoc
\usepackage[caption=false,font=normalsize,labelfont=sf,textfont=sf]{subfig}
\else
\usepackage[caption=false,font=footnotesize]{subfig}
\fi

\begin{document}

\title{Design and Evaluation of a Novel Short Prototype Filter for FBMC/OQAM Modulation}

\author{ Jeremy~Nadal \IEEEmembership{Student Member, IEEE}, 
              Charbel~Abdel~Nour \IEEEmembership{Member, IEEE}, 
							Amer~Baghdadi \IEEEmembership{Senior Member, IEEE}\\

\thanks{J. Nadal, C. Abdel Nour and A. Baghdadi are with IMT Atlantique, CNRS Lab-STICC, UBL, F-29238 Brest, France (e-mail:\{firstname.lastname\}@imt-atlantique.fr)}

}

\maketitle

\begin{abstract}
Filter-Bank Multi-Carrier with Offset Quadrature Amplitude Modulation (FBMC/OQAM) is considered by recent research projects as one of the key enablers for the future 5G air interface. 
It exhibits better spectral shape and improves mobility support compared to Orthogonal Frequency-Division Multiplexing (OFDM) thanks to the use of a time and frequency localized prototype filter. 
The choice of this filter is crucial for FBMC/OQAM, due to its substantial impact on achieved performance and complexity levels. 
In the context of 5G, short frame sizes are foreseen in several communication scenarios to reduce system latency, and therefore short filters are preferred. 
In this context, a novel short filter allowing for near perfect reconstruction and having the same size as one OFDM symbol is proposed. 
Using Frequency-Spread (FS) implementation for the FBMC/OQAM receiver, analytical analysis and simulation results show that the proposed filter exhibits better robustness to several types of channel impairments when compared to State-of-The-Art (SoTA) prototype filters and OFDM modulation. 
In addition, FS-based hardware architecture of the filtering stage is proposed, showing lower complexity than the classical PolyPhase Network (PPN)-based implementation. 
\end{abstract}


%
\IEEEpeerreviewmaketitle

{}

\section{Introduction}

Next generation mobile communication systems are foreseen to provide ubiquitous connectivity and seamless service delivery in all circumstances. 
The expected important number of devices and the coexistence of human-centric and machine-type applications will lead to a large diversity of communication scenarios and characteristics \cite{5GPP2016}. 
In this context, many advanced communication techniques are under investigation. 
Taken individually, each one of these techniques is suitable for a subset of the foreseen communication scenarios.

Filter-Bank Multi-Carrier with Offset Quadrature Amplitude Modulation (FBMC/OQAM), or in short FBMC, is being studied and considered nowadays by recent research projects as a key enabler for the future flexible 5G air interface \cite{SZL+2014}. 
It exhibits better spectrum shape compared to the traditional Orthogonal Frequency-Division Multiplexing (OFDM) and enables better spectrum usage and improved mobility support. 
This is possible thanks to the use of a Prototype Filter (PF) which allows to improve the time and/or frequency localization properties of the transceiver. 
The orthogonality is preserved in the real domain (as opposite to complex domain) with the OQAM scheme. 
Furthermore, FBMC implementation relies on Fast Fourier Transform (FFT), similarly to OFDM, with an additional low-complexity PolyPhase Network (PPN) filtering stage.

However, the choice of the PF is crucial for the design of an efficient FBMC modulation. 
In fact, the time/frequency localization of this filter can impact significantly the different performance levels \cite{LGS2014} and the frame structure of the communication system. 
Furthermore, the length of the PF impacts considerably the transceiver complexity. 
Thus, the careful design of new PFs is of high interest to improve robustness of FBMC against channel impairments and to support the constraints imposed by various 5G scenarios. 

In this context, a novel short PF is proposed. It is obtained by inverting the time and frequency lattice of the Filter-Bank (FB) impulse response of a long PF. 
Due to its near perfect reconstruction property and its length of one OFDM symbol, it is denoted by Near Perfect Reconstruction 1 (NPR1) in this paper. 

In-depth technical analysis and comparison with existing short PFs in terms of power spectral density, robustness to timing and frequency offsets, and robustness to multipath channel impairments for different wireless channel models are performed. 
Both PPN and Frequency Spread (FS) implementations of the FBMC receiver, respectively referred to as PPN-FBMC and FS-FBMC in this paper, are considered. It is shown that:

\begin{enumerate}
\item the NPR1 PF achieves improved robustness against all the considered channel impairments when the FS-FBMC receiver is considered,
\item the FS-FBMC receiver offers improved robustness against timing offset and multipath impairments when compared to the PPN-FBMC receiver,
\item by exploiting the different properties of the PFs, a substantial reduction in hardware complexity can be achieved for the FS-FBMC receiver. 
\end{enumerate}

Particularly, it is shown in this paper that the hardware complexity of the FS-FBMC receiver is lower than the PPN-FBMC receiver when the NPR1 PF is considered. 

The rest of the paper is organized as follows. 
Section \ref{Section_FBMC_OQAM_System_Description} provides a technical description of the FBMC modulation with different types of implementation. 
Section \ref{Section_SoTA_on_short_prototype_filter_and_proposal} is dedicated to the presentation of the proposed novel short PF along with State-of-The-Art (SoTA) existing ones. 
Section \ref{Section_Performance_evaluation_of_short_filter} evaluates and compares the performance of all considered PFs with several channel impairments. 
Section \ref{section_complexity_eval} presents the proposed hardware architecture of the FS filtering stage and illustrates the complexity reduction with respect to the PPN-FBMC architecture.  
Finally, Section \ref{Section_Conclusion} concludes the paper.


\section{FBMC/OQAM system description}
\label{Section_FBMC_OQAM_System_Description}

\begin{figure*}[!t]
\centering
\normalsize
\includegraphics[width=6in]{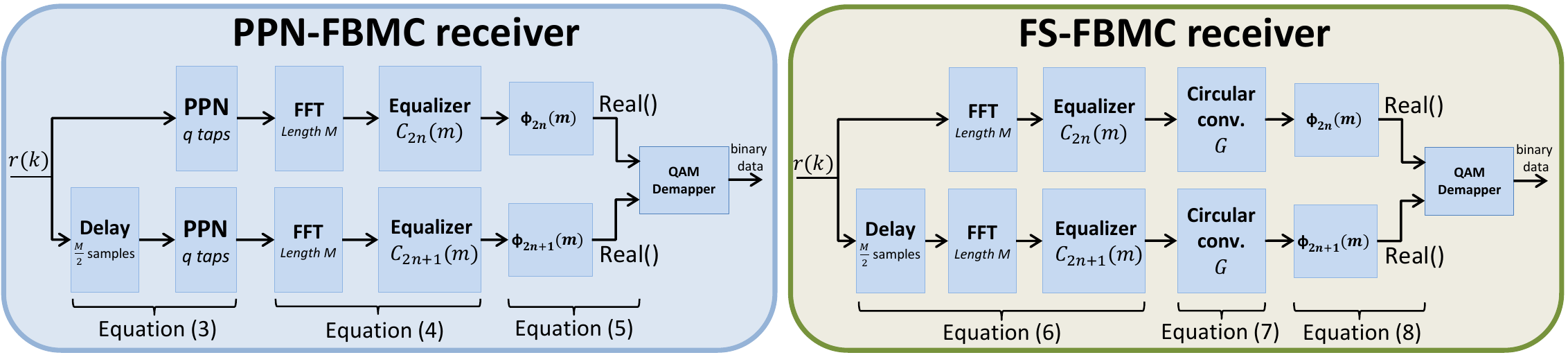}
\caption{System description of the PPN and FS implementations of the FBMC modulator.}
\label{figure_PPN_FS_system}
\end{figure*}

FBMC is a multicarrier transmission scheme that introduces a filter-bank to enable efficient pulse shaping for the signal conveyed on each individual subcarrier. 
This additional element represents an array of band-pass filters that separate the input signal into multiple components or subcarriers, each one carrying a single frequency sub-band of the original signal.  
As a promising variant of filtered modulation schemes, FBMC, originally proposed in \cite{S1967} and also called OFDM/OQAM \cite{SSL2002} or staggered modulated multitone (SMT) \cite{SSO2015}, can potentially achieve a higher spectral efficiency than OFDM since it does not require the insertion of a Cyclic-Prefix (CP).
Additional advantages include the robustness against highly variant fading channel conditions and imperfect synchronization by selecting the appropriate PF type and coefficients \cite{LGS2014}. 
Such a transceiver structure usually requires a higher implementation complexity related not only to the filtering steps but also to the applied modifications to the modulator/demodulator architecture. 
However, the usage of digital polyphase filter bank structures \cite{BD1974}\cite{SSL2002}, together with the rapid growth of digital processing capabilities in recent years have made FBMC a practically feasible approach.

In the literature, two types of implementation for the FBMC modulation exist, each having different hardware complexity and performance. The first one is the PPN implementation \cite{H1981}, illustrated in Figure \ref{figure_PPN_FS_system}, which is based on an IFFT and a PPN for the filtering stage, and enables a low complexity implementation of the FBMC transceiver.  

The second type of implementation is the FS implementation (Figure \ref{figure_PPN_FS_system}), proposed in \cite{B2012a}\cite{M1998} for the Martin--Mirabassi--Bellange PF with an overlapping factor equal to 4  (MMB4), considered for FBMC during PHYDYAS project \cite{PHYD}. 
The original idea was to shift the filtering stage into the frequency domain, in order to enable the use of a low-complexity per-sub-carrier equalizer as in OFDM. 
The hardware complexity is supposed to be higher than the complexity of the PPN implementation, at least for long PFs. 
In fact, it requires one FFT of size $L = KM$ per FBMC symbol, where $K$ is the overlapping factor of the PF, and $M$ is the total number of available sub-carriers. 
However, in the short PF case ($K=1$), the size of the FFT is same as for the PPN implementation.

The rest of the section provides a mathematical background of the PPN-FBMC transceiver and the FS-FBMC receiver.

\subsection{PolyPhase Network-based implementation}

If $M$ is the total number of available sub-carriers and $a_n(m)$ the Pulse-Amplitude Modulation (PAM) symbol at subcarrier index $m$ and time slot $n$, then the baseband signal $s(k)$ can be mathematically decomposed as follows:
\setlength{\arraycolsep}{0.0em}
\begin{eqnarray}
\label{equation_sk}
&& s(k) {}={}\sum_{n=-\infty}^{+\infty} g\bigl( k-n\frac{M}{2} \bigr) \, x_n(k),  \\
\label{equation_xnk}
&& x_n(k){}={} \sum_{m=0}^{M-1} (-1)^{nm} a_n(m) \phi_n(m) e^{i2\pi \frac{km}{M}},
\end{eqnarray}
\setlength{\arraycolsep}{5pt}
with $i^2 = -1$. To keep the orthogonality in the real field, $\phi_n(m)$ must be a quadrature phase rotation term. 
In the literature, it is generally defined as $\phi_n(m) = i^{n+m}$, as in \cite{LGS2014}.
The impulse response of the PF is $g$, with $g(l) = 0$ when $l \not\in [0,L-1]$, where $L$ is the length of the PF. 
In practice, the PPN-FBMC transmitter is implemented using an IFFT of size $M$ followed by a PolyPhase Network. 
When a short PF is used ($K=1$), this latter can be seen as a windowing operation: the outputs of the IFFT are simply multiplied by the PF impulse response $g$. 
Consequently, the complexity overhead introduced by the PPN is limited. 
Note that, due to the OQAM scheme, the obtained FBMC symbol overlaps with both the previous and next symbols on half of the symbol length. 
Therefore, for practical implementation, $2$ FBMC symbols may be generated in parallel. 
It is however possible to avoid the use of two IFFT blocks at the transmitter side through the use of the pruned FFT algorithm. 
This leads to a reduced-complexity implementation presented in \cite{D2011} and \cite{NNB2015}. 

Receiver side implementation applies dual operations with respect to the ones performed by the constituent blocks of the transmitter. 
The IFFT must be replaced by an FFT, and the operations order must be reversed: PPN (windowing if $K=1$), FFT then OQAM demapper, as shown in Figure~\ref{figure_PPN_FS_system}. 
If $r$ is the received signal and $\hat{a}_n(m)$ are the recovered PAM symbols, then we have:
\setlength{\arraycolsep}{0.0em}
\begin{eqnarray}
\label{equation_unm}
&& u_n(m) {}={} \sum_{l=0}^{K-1} g(k+lM) \, r \biggl( k+\frac{M}{2}(2l+n) \biggr). \\
\label{equation_Unm}
&& U_n(m) {}={} C_n(m) \sum_{k=0}^{M-1} u_n(m) e^{i2\pi \frac{km}{M} }. \\
&& \hat{a}_n(m) {}={} \Re \Bigl( \phi^*_n(m) U_n(m) \Bigr).
\label{equation_rec_amn_PPN}
\end{eqnarray}
\setlength{\arraycolsep}{5pt}
where $*$ represents the complex conjugate operation, and $C_n(m)$ is the Zero Forcing (ZF) equalizer coefficient to compensate the impairments introduced by the channel. 
Note that, contrary to transmitter side, doubling the FFT processing cannot be avoided using the pruned FFT algorithm. 
The main reason is that the equalization term $C_n(m)$ introduces complex valued coefficients.

\subsection{FS-FBMC receiver description}

The FS-FBMC implementation is generally considered at the receiver side, since it enables a low-complexity and efficient equalization scheme \cite{B2012b}\cite{BDN2014}. 
It remains perfectly compatible with the PPN implementation at the transmitter side. The received symbols are expressed as follows:
\setlength{\arraycolsep}{0.0em}
\begin{eqnarray}
\label{equation_Xnm}
&& X_n(m){}={} C_n(m) \sum_{k=0}^{L-1} r(k+n\frac{M}{2})e^{-i\frac{2\pi }{L}km}  \\
\label{equation_Ynm}
&& Y_n(m){}={} \sum_{l=-\frac{L}{2}}^{\frac{L}{2}-1} G(l) X_n(m-l)\\
\label{equation_amn_recovered}
&& \hat{a}_{n}(m){}={} \Re \Bigl( Y_n(Km) \phi^*_n(m) \Bigr)
\end{eqnarray}
\setlength{\arraycolsep}{5pt}
where $G$ is the frequency response of the PF. 
The FS-FBMC receiver first applies an IFFT of size $L$ on the part of the $r$ signal containing the FBMC symbol to demodulate in order to obtain the $X_n$ signal in frequency domain (\ref{equation_Xnm}). 
Then, it introduces a filtering stage in frequency domain, as described in (\ref{equation_Ynm}): the frequency response of the PF $G$ is convoluted with the $X_n$ signal, for instance using a Finite Impulse Response (FIR) filter. Finally, the recovered PAM symbols are obtained by extracting the real part of quadrature phase rotated and down-sampled $Y_n$ samples (\ref{equation_amn_recovered}).

These operations are summarized and illustrated in Figure~\ref{figure_PPN_FS_system}. 
The FS-FBMC implementation seems highly complex, however $G$ has a lot of zero coefficients due to its frequency localization. 
Therefore, it can be truncated down to $N_G$ coefficients. Then, by defining $\Delta = (N_G - 1)/2$ ($N_G$ is considered an odd number), (\ref{equation_Ynm}) becomes:
\setlength{\arraycolsep}{0.0em}
\begin{eqnarray}
&& Y_n(m)(k){}={} \sum_{l=-\Delta}^{\Delta -1} G(l) X_n(m-l)\\  \nonumber
\end{eqnarray}
\setlength{\arraycolsep}{5pt}

\section{Proposal of a novel short prototype filter}
\label{Section_SoTA_on_short_prototype_filter_and_proposal}

Current literature often focuses on FBMC using a PF with a duration $4$ times larger than an OFDM symbol ($K = 4$), like MMB4 or Isotropic Orthogonal Transform Algorithm 4 (IOTA4) \cite{FAB1995}. 
However, a shorter PF can also be applied, as proposed in \cite{PSS2004} with the Time-Frequency Localization 1 (TFL1) PF. 
Another example is the Quadrature Mirror Filter 1 (QMF1) \cite{M1990} which was recently applied to FBMC leading to a variant denoted by Lapped-OFDM modulation and presented in \cite{BMT2015}. 
In the rest of this paper, PFs with a duration larger than one OFDM symbol will be referred to as long PFs (e.g. MMB4, IOTA4), and the ones with a duration of one OFDM symbol as short PFs (e.g. TFL1, QMF1). 
When compared to long PFs, short PFs provide lower latency, higher robustness against Carrier Frequency Offset (CFO) and higher spectral efficiency due to shortened transition between two successive transmission frames. 
In this context, finding a new short PF with good performance and low hardware complexity becomes a challenging task of high interest. 
In the following, after a short description of the two existing short PFs in the literature, we present a novel short PF design that shows significant advantages in terms of performance and complexity.

\begin{table}[!t]

\caption{$X_i$ coefficients of the analytical expression of TFL1 PF.}
\centering
\begin{tabular}{|c|c|c|c|c|c|c|}

\hline
$l$   & $X_l$                 & $l$ & $X_l$\\
\hline
$0$   & $4.1284847578$        & $4$ & $-2.1107642825\,10^1$  \\
$1$   & $1.9727736832$        & $5$ & $-6.6774831778\,10^{-3}$ \\
$2$   & $1.2781855004\,10^{-1}$ & $6$ & $-1.0150558822\,10^2$  \\
$3$   & $-1.4505800309\,10^2$ & $7$ & $1.9143799092\,10^{-2}$  \\
\hline

\end{tabular}
\label{table_Xi_coefficients}
\end{table}

\subsection{TFL1 and QMF1 prototype filters}

The TFL1 PF was the first attempt to specifically design a time and frequency localized short PF for the FBMC modulation \cite{PSS2004}. 
It is the most known short PF in the literature.
Indeed, it is already integrated into proof-of-concept hardware platforms \cite{LJG+2012}\cite{NAB+2014}. 
The analytical expression of the TFL1 PF \cite{PS2013} is given by: 
\setlength{\arraycolsep}{0.0em}
\begin{eqnarray}
&&g(k){}={} \frac{\pi}{2}(1-x) + \gamma_0 t + 2t(t^2 - 1)(\beta_1 + 4 \beta_2 t^2), \nonumber
\label{eq_FBMC_decomposed}
\end{eqnarray}
\setlength{\arraycolsep}{5pt}
where $x = \frac{2k+1}{M}$ with $k \in [0,\frac{M}{2}-1]$, $t = 2x-1$ and
\setlength{\arraycolsep}{0.0em}
\begin{eqnarray}
&&\gamma_0(k){}={} \frac{1}{X_0 + X_1\frac{M}{2}} \nonumber \\ 
&&\beta_1(k){}={} X_2 + \frac{1}{X_3 + X_4\frac{M}{2}} \nonumber \\
&&\beta_2(k){}={} X_5 + \frac{1}{X_6 + X_7\frac{M}{2}},\nonumber
\end{eqnarray}
\setlength{\arraycolsep}{5pt}
$X_l$ being defined in Table \ref{table_Xi_coefficients}. The second half of the PF coefficients are constructed by symmetry: $g(k) = g(M-k-1)$ for $k \in [\frac{M}{2},M-1]$.

Regarding the QMF1 PF \cite{M1990}, it was applied to FBMC leading to a variant denoted by Lapped-OFDM modulation presented in \cite{BMT2015}. The analytical expression of the QMF1 PF is given below:
\setlength{\arraycolsep}{0.0em}
\begin{eqnarray}
g(k) = \text{sin}\left(\frac{ \pi k } { M } \right).
\end{eqnarray}
\setlength{\arraycolsep}{5pt}
\begin{table}[!t]
\caption{Filter-bank impulse response of the MMB4 filter.}
\centering
\begin{tabular}{|c|c|c|c|}
\hline
  \backslashbox{q}{p} & $-1$       & $0$      & $1$ 	     \\
\hline
$-\frac{3M}{2}$ & $i\,0.043$  & $-0.067$ & $-i\,0.043$ \\
\hline
$-M$ & $-0.125$    & $0$      & $-0.125$     \\
\hline
$-\frac{M}{2}$ & $-i\,0.206$ & $0.564$  & $i\,0.206$  \\
\hline
$0$   & $0.239$     & $1$      & $0.239$     \\
\hline
$\frac{M}{2}$ & $i\,0.206$  & $0.564$  & $-i\,0.206$     \\
\hline
$M$ & $ -0.125$   & $0$      & $-0.125$  \\
\hline
$\frac{3M}{2}$ &$ -i\,0.043$ & $-0.067$ & $i\,0.043$  \\
\hline
\end{tabular}
\label{table_impulse_response_MMB4}
\end{table}

\subsection{Proposed Near Perfect Reconstruction filter}
\label{subsection_proposed_NPR1_filter}

This sub-section describes a novel short PF representing one of the major contributions of this paper. 
The main design procedure of the proposed PF starts by inverting the time and frequency axes of the Filter-Bank (FB) impulse response of the MMB4 PF. 
The coefficients of this FB impulse response are given in \cite{Bel2010} and presented in Table \ref{table_impulse_response_MMB4}. 
It can be seen that the FB impulse response of the MMB4 PF is highly localized in frequency since interference is limited only to one adjacent sub-carrier (indexes $p=-1$ and $p=1$) in the frequency plane. 
Inverting the time ($q$) and frequency ($p$) axes will generate a PF highly localized in time, since the obtained FB impulse response coefficients have values only at the adjacent FBMC symbols ($q = -M/2$ and $q = M/2$). 
Therefore, a PF with an overlapping factor of $1$ is sufficient to obtain these coefficients. 
Consequently, the PF coefficients can be deduced from a given FB impulse response.

By definition, the FB impulse response is composed of the values obtained at the output of the receiver $Y_n(m)$ from (\ref{equation_Ynm}) by setting $a_{0}(0) = 1$ and $a_n(m) = 0$ when $(n,m) \neq (0,0)$. In this case, we have $r(k) = g(k)$, and $U_n(m)$ in (\ref{equation_Unm}) becomes:
\setlength{\arraycolsep}{0.0em}
\begin{eqnarray}
U_n(m) = \sum_{k=0}^{M-1} g(k) g(k+n\frac{M}{2}) \, e^{-i\frac{2\pi km}{M}}. 
\end{eqnarray}
\setlength{\arraycolsep}{5pt}
\noindent Furthermore, we have $U_n(m) = F_{g_{MMB4}}(n,m)$, where $F_{g_{MMB4}}(p,q)$ is the FB impulse response coefficients of the MMB4 PF presented in Table \ref{table_impulse_response_MMB4}. Particularly, for $n = 0$, we have:
\setlength{\arraycolsep}{0.0em}
\begin{eqnarray}
U_0(m) = \sum_{k=0}^{M-1} g(k)^2 \, e^{-i\frac{2\pi km}{M}}. 
\end{eqnarray}
\setlength{\arraycolsep}{5pt}
Thus, $g$ can be deduced as follows:
\setlength{\arraycolsep}{0.0em}
\begin{eqnarray}
g(k) = \sqrt{ \sum_{l=0}^{M-1} F_{g_{MMB4}}(0,l)\,e^{i\frac{2\pi kl}{M}} }
\end{eqnarray}
\setlength{\arraycolsep}{5pt}
Then, the design procedure introduces simplifications to obtain a simpler analytical expression, by taking advantage of the real valued and symmetrical $F_{g_{MMB4}}$ coefficients:
\setlength{\arraycolsep}{0.0em}
\begin{eqnarray}
&&g(k) = \sqrt{ 1 - 2\,\sum_{l=0}^{2} P_g(l) \text{cos}\Bigl( \frac{2\pi k(2l+1)}{M} \Bigr) }\\
&&P_g(0) =  0.564447 \nonumber\\
&&P_g(1) = -0.066754 \nonumber\\
&&P_g(2) =  0.002300 \nonumber
\label{equation_g_NPR1}
\end{eqnarray}
\setlength{\arraycolsep}{5pt}
We call the resulting proposed short PF with overlapping factor equal to $1$ as Near Perfect Reconstruction (NPR1) PF due to its nature, similar to the MMB4 PF. 
To analytically calculate the residual interference, the recovered PAM symbols $\hat{a}_n(m)$ must be expressed by taking into account the transmitted ${a}_n(m)$ symbols and the effect of the PF at the transmitter. 
Thus, by setting $r(k) = s(k)$ and by integrating equations (\ref{equation_sk}), (\ref{equation_unm}) and (\ref{equation_Unm}) into (\ref{equation_rec_amn_PPN}), we have:
\setlength{\arraycolsep}{0.0em}
\begin{eqnarray}
\hat{a}_n(m) && =  \Re \Biggl[ \phi^*_n(m) \sum_{k=0}^{M-1} g(k) \\ 
&& \times \sum_{n'=-\infty}^{+\infty}  g \Bigl( k+(n-n')\frac{M}{2} \Bigr) x_{n'}\Bigl(k+n\frac{M}{2}\Bigr) e^{-i2 \pi \frac{km}{M}} \Biggr] \nonumber
\label{equation_recPAM}
\end{eqnarray}
\setlength{\arraycolsep}{5pt}
\noindent Due to the time localization of the PF, we have $(n-n') = q \in \llbracket -Q,Q \rrbracket$, $Q$ being the number of FBMC symbols acting as interference after (or before) the FBMC symbol currently demodulated. 
Typically, we have $Q = 1$ for short PFs, since the impulse response of $g$ is equal to zero after $M$ samples. Then, we have: 
\setlength{\arraycolsep}{0.0em}
\begin{eqnarray}
\label{equation_recPAM2}
\hat{a}_n(m) && =  \Re \Biggl[ \phi^*_n(m) \sum_{q=-Q}^{Q} \sum_{k=0}^{M-1} g(k) g \Bigl( k+q\frac{M}{2} \Bigr) \nonumber \\ 
             && \,\,\,\,\, \times \, z_{n-q}(k) e^{-i2 \pi \frac{km}{M}} \Biggr]  \\
\end{eqnarray}
\setlength{\arraycolsep}{5pt}
with
\setlength{\arraycolsep}{0.0em}
\begin{eqnarray}
z_n(k) && =  x_n \Bigl( k + n\frac{M}{2} \Bigr) \nonumber \\ 
       && = \sum_{m=0}^{M-1} a_n(m) \phi_n(m) e^{i2\pi\frac{km}{M}}. \nonumber
\end{eqnarray}
\setlength{\arraycolsep}{5pt}
In (\ref{equation_recPAM2}), the term corresponding to the FFT of  $g(k) g \Bigl( k+qM/2 \Bigr) \times z_{n-q}(k)$ can be rewritten by a circular convolution operation denoted by $\circledast$, as follows: 
\setlength{\arraycolsep}{0.0em}
\begin{eqnarray}
\hat{a}_n(m) && =  \Re \Biggl[ \phi^*_n(m) \sum_{q=-Q}^{Q} \Bigl( F_g(m,q) \circledast Z_{n-q}(m) \Bigr) \Biggr],  \nonumber \\
\end{eqnarray}
\setlength{\arraycolsep}{5pt}
with 
\setlength{\arraycolsep}{0.0em}
\begin{eqnarray}
\Bigl( F_g(m,q) \circledast Z_{n-q}(m) \Bigr) && = \sum_{p=-\frac{M}{2}}^{\frac{M}{2}-1} F_g(p,q\frac{M}{2}) Z_{n-q}(m-p),  \nonumber \\
\end{eqnarray}
\setlength{\arraycolsep}{5pt}
\noindent where $F_g(p,q)$ and $Z_n(m-p)$ are the results of the application of a FFT to the terms $g(k) g(k+q)$ and $z_{n-q}(k)$, respectively expressed as:
\setlength{\arraycolsep}{0.0em}
\begin{eqnarray}
F_{g}(p,q) && = \sum_{k=0}^{M-1}g(k)g(k+q) e^{-i2 \pi \frac{pk}{M}},
\end{eqnarray}
\setlength{\arraycolsep}{5pt}
and,
\setlength{\arraycolsep}{0.0em}
\begin{eqnarray}
Z_n(m) && = a_n(m)\phi_n(m).\nonumber
\end{eqnarray}
\setlength{\arraycolsep}{5pt}
\noindent Finally, the expression of the recovered PAM symbol $\hat{a}_n(m)$ becomes: 
\setlength{\arraycolsep}{0.0em}
\begin{eqnarray}
\hat{a}_n(m) && = \Re \Biggl[ \sum_{(p,q) \in \Omega} \phi^*_n(m)\phi_{n-q}(m-p) F_g(p,q\frac{M}{2}) \nonumber \\ 
             && \,\,\,\,\, \times \, a_{n-q}(m-p) \Biggr], \nonumber \\
             && = \Re \Biggl[ \sum_{(p,q) \in \Omega} i^{-q-p} F_g(p,q\frac{M}{2}) a_{n-q}(m-p) \Biggr]
\end{eqnarray}
\setlength{\arraycolsep}{5pt}
with $\Omega = \llbracket -M/2,M/2-1 \rrbracket \times \llbracket -Q,Q \rrbracket$. 
In fact, $F_g(p,q)$ is the FB impulse response of $g$, and for the NPR1 PF, we have $F_{g_{NPR1}}(p,q) = F_{g_{MMB4}}(q,p)$. 
If $a_n(m)$ symbols are independent and identically distributed random variables and $ \mathbb{E}(a_n(m)) = 0$, then the residual interference of the PF can be evaluated as follows:
\setlength{\arraycolsep}{0.0em}
\begin{eqnarray}
\text{SIR}_\text{NPR1} && = \frac{ \mathbb{E} \Bigl( \Re \Bigl( F_g(0,0) a_n(m) \Bigr)^2 \Bigr) }{ \mathbb{E}  \Bigl( (F_g(0,0)a_n(m) - \hat{a}_n(m))^2 \Bigr) } \nonumber \\
           && = \frac{ \Re \Bigl( F_g(0,0) \Bigr)^2 }{ \sum_{(p,q) \in \Omega^0} \Re \Bigl[ i^{p+q} F_g(p,q\frac{M}{2}) \Bigr]^2  } \label{equation_SIR_NPR}
\end{eqnarray}
\setlength{\arraycolsep}{5pt}
\noindent with SIR being the Signal-to-Interference Ratio and $\Omega^0 = \Omega \setminus \{(0,0)\}$. 
For $M = 512$, and using the coefficients presented in \cite{M1998} to design the MMB4 PF and the related FB impulse response, we have $F_g(0,0) = 1$ and the obtained SIR is $73$ dB for the proposed NPR1 PF.
This SIR has the same order of magnitude as the SIR of the MMB4 PF ($\approx 70$ dB \cite{Bel2010}\cite{M1998}), confirming the near perfect reconstruction nature of the proposed short filter.

\section{Performance evaluation}
\label{Section_Performance_evaluation_of_short_filter}

This section evaluates and compares the performance of the proposed NPR1 short PF with respect to SoTA ones. 
It provides a comparison of different FBMC short PFs, including the proposed one, in terms of spectral usage and SIR when applying a truncated FS implementation. 
Their robustness against several types of channel impairments is also evaluated and compared with OFDM, for both PPN and FS implementations. 
These impairments include timing synchronization errors, carrier frequency offset and the use of a multipath channel.

\begin{figure}[!t]
\centering
\normalsize
\includegraphics[width=3in]{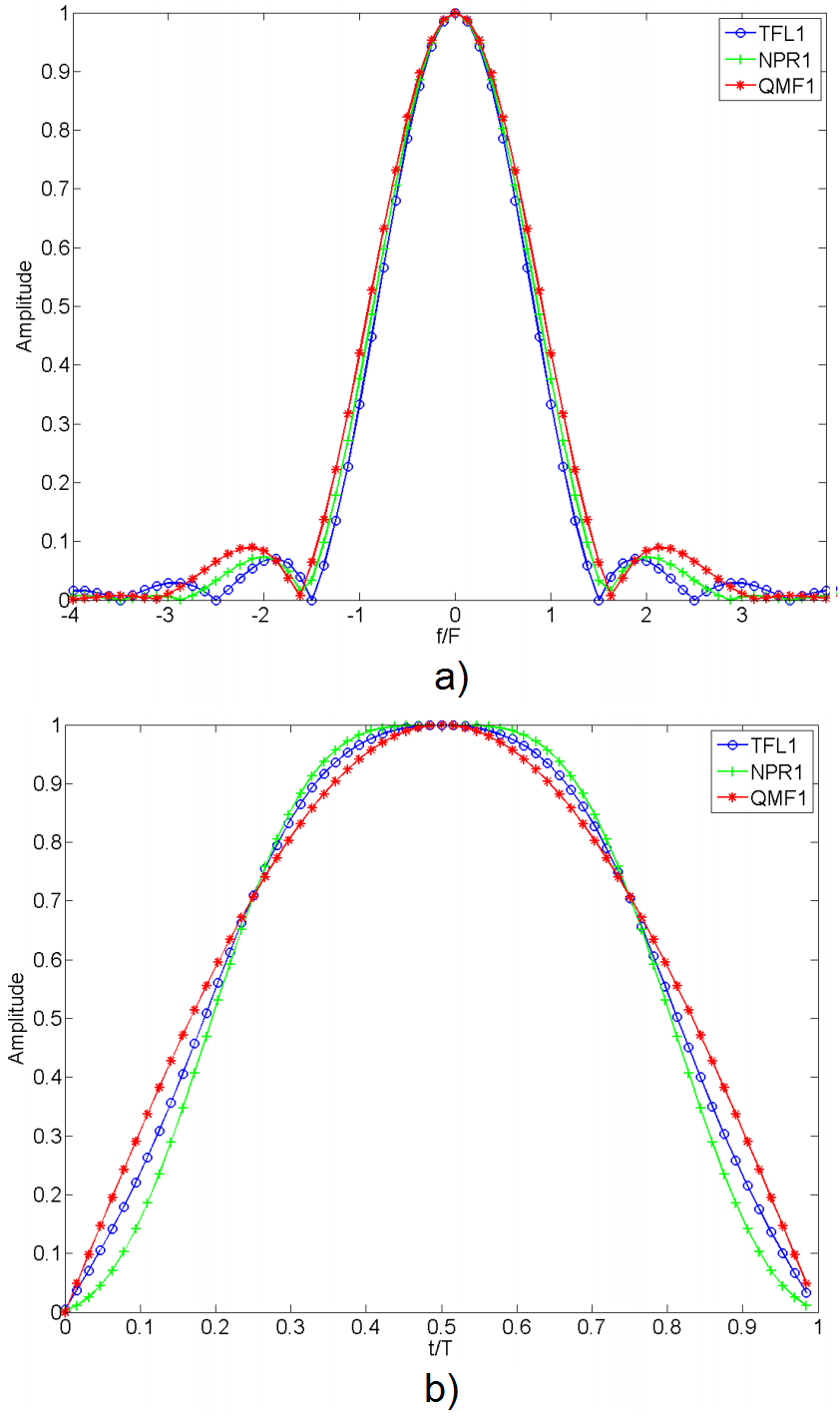}
\caption{Frequency response a) and impulse response b) of the TFL1, QMF1 and NPR1 PFs.}
\label{figure_IR_FS_shape}
\end{figure}

\subsection{Comparison of out-of-band power leakage}

One of the main advantages of FBMC over OFDM resides in its spectrum shape with low Out-of-Band Power Leakage (OOBPL). 
Consequently, a shorter guard-band can be used to fit the Adjacent Channel Leakage Ratio (ACLR) constraints and to support relaxed synchronization communication services. 
In general, long PFs have lower OOBPL when compared to short PFs on one side, but lose the other advantages of short PFs provided in Section \ref{Section_SoTA_on_short_prototype_filter_and_proposal} on the other side.

\begin{figure}[!t]
\centering
\normalsize
\includegraphics[width=3in]{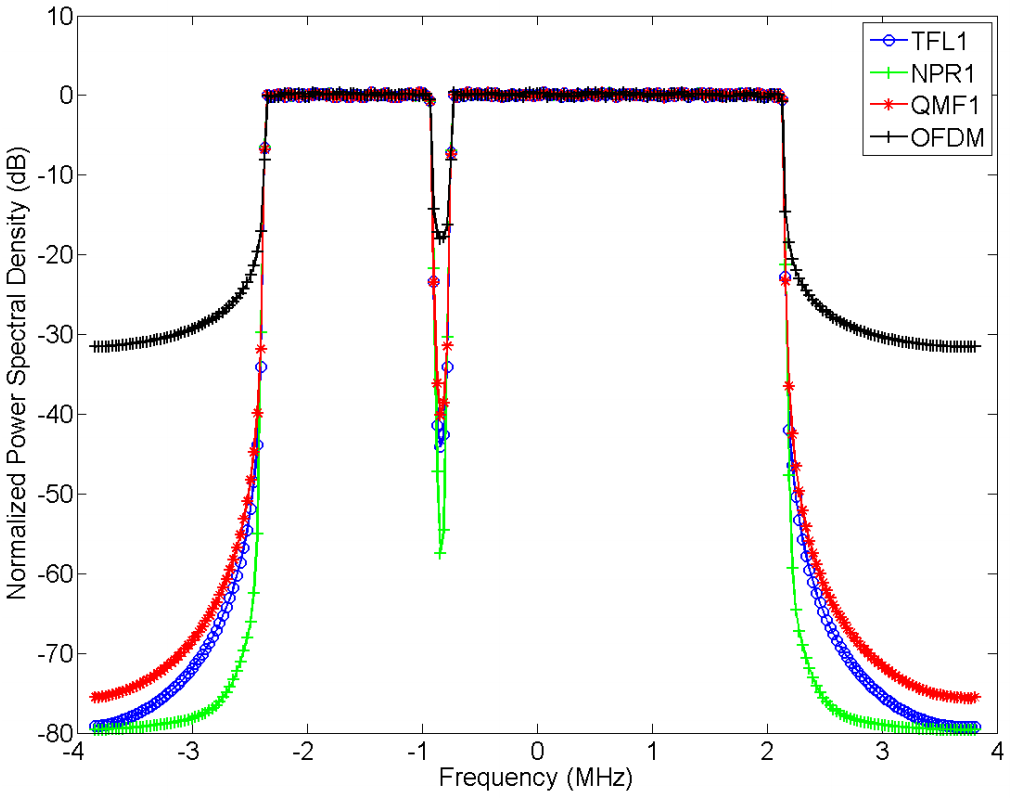}
\caption{PSD evaluation of the FBMC short PFs in a 4.5 MHz bandwidth.}
\label{figure_PSD_eval}
\end{figure}

However, depending on the chosen short PF, the spectral characteristics may vary. The frequency response of the QMF1 PF, depicted in Figure \ref{figure_IR_FS_shape}a, has the secondary lobes with the highest amplitude, followed by TFL1 then NPR1 PF. This explains the reasons behind the differences in the OOBPL depicted in Figure \ref{figure_PSD_eval}. Indeed, this figure shows the power spectral density of OFDM and FBMC with different short PFs, on a $4.5$ MHz bandwidth. Simulation parameters correspond to a 4G/LTE setting were a notch of $12$ subcarriers, or 1 Ressource Block (RB), was inserted in the spectrum to evalute the capacity to support fragmented spectrum for asynchronous communication services.

As expected, the OOBPL is extremely low for FBMC: a gap of $59$ dB can be observed between OFDM and FBMC at the extreme edges of the bandwidth, independently from the used PF. For NPR1 case, the OOPBL quickly decreases when compared to the other PFs, since it has the lowest secondary lobes (Figure \ref{figure_IR_FS_shape}a). Inside the notch, a gap of $40$ dB can be observed between OFDM and FBMC with NPR1 PF, and a difference of $17$ dB between this PF and QMF1. These results demonstrate that, despite using a short PF, the OOBPL is still very low for FBMC when compared to OFDM, even in a fragmented band. In conclusion, NPR1 represents the most suitable short PF to respect high ACLR constraints.

\subsection{Truncation impact on the frequency response of the filter}
\label{subsec_truncated_impact}

The frequency response of the PF can be truncated to reduce the complexity of the FS-FBMC receiver. By truncating the frequency response of the PF at the receiver side, interference may appear due to a non perfect reconstruction, resulting in performance degradation. 
However, if $N_G$, the number of non-zero coefficients, is too high, the resulting FS implementation will require important hardware complexity. 
A compromise between complexity and performance must be devised. 

\begin{figure}[!t]
\centering
\normalsize
\includegraphics[width=3.5in]{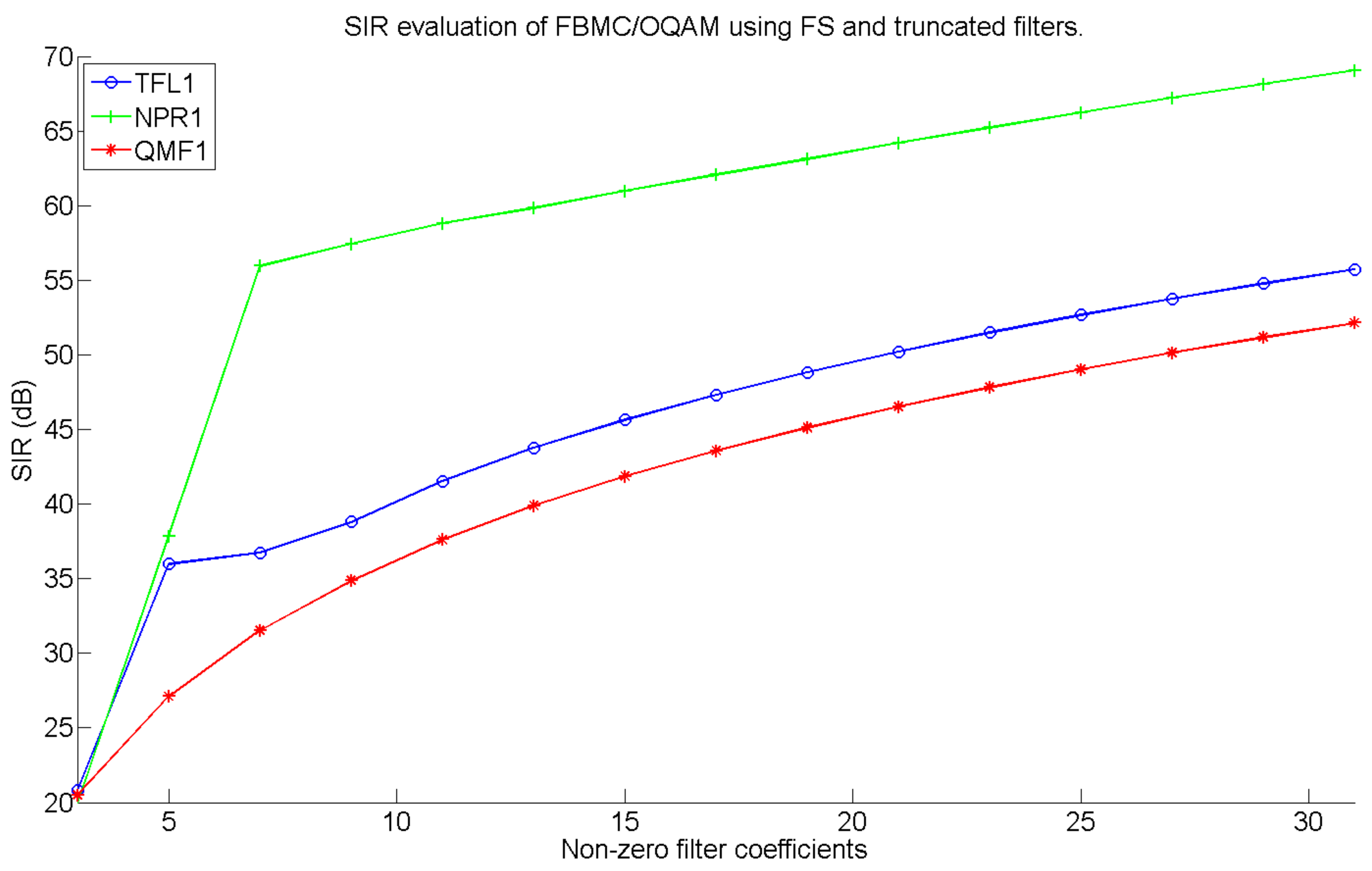}
\caption{Evaluation of the impact of $N_G$ (number of non-zero filter coefficients) on the SIR for different PFs.}
\label{figure_FS_truncation}
\end{figure}

Equation (\ref{equation_SIR_NPR}) can be adapted to evaluate the residual interference introduced by the truncation. In this case, the PF impulse response $g$ is replaced by the truncated one $\tilde{g}$ in (\ref{equation_recPAM}), where $\tilde{g}$ is expressed as follows:
\begin{equation}
\tilde{g}(k) = \sum_{l=-\Delta}^{\Delta} G(l) e^{i2 \pi \frac{kl}{M}}.
\end{equation}
The values of $\tilde{g}$ are obtained by computing an IFFT of size $M$ on the $N_G$ non-zero coefficients of $G$. 
Then, using similar mathematical development as described in Sub-section \ref{subsection_proposed_NPR1_filter} from (\ref{equation_recPAM}), the analytic expression of the SIR is:
\setlength{\arraycolsep}{0.0em}
\begin{eqnarray}
\text{SIR}_{\text{trunc}} && = \frac{ \Re \Bigl( F_{g,\tilde{g}}(0,0) \Bigr)^2 }{ \sum_{(p,q) \in \Omega^0} \Re \Bigl[ i^{p+q} F_{g,\tilde{g}}(p,q\frac{M}{2}) \Bigr]^2  }
\label{equation_SIR_trunc}
\end{eqnarray}
\setlength{\arraycolsep}{5pt}
where $F_{g,\tilde{g}}(p,q)$ is the FB impulse response using the PF $g$ at the transmitter side and the PF $\tilde{g}$ at the receiver side. It is expressed as follows:
\setlength{\arraycolsep}{0.0em}
\begin{eqnarray}
F_{g,\tilde{g}}(p,q) && = \sum_{k=0}^{M-1}g(k+q)\tilde{g}(k) e^{-i2 \pi \frac{pk}{M}}.
\end{eqnarray}
\setlength{\arraycolsep}{5pt}
The obtained numerical values of the SIR are presented in Figure \ref{figure_FS_truncation} for different PFs and corresponding $N_G$ values. 
The analytical results of (\ref{equation_SIR_trunc}) have also been confirmed by simulations.
Table \ref{table_truncation_SIR} summarizes the needed number of non-zero coefficients for a SIR target ranging between $50$ and $70$ dB, depending on the used PF. 

\begin{table}[!b]
\caption{$N_G$ values needed to reach target SIR for different PFs.}
\centering
\begin{tabular}{|c|c|c|c|c|c|c|}
\hline
SIR (dB) & TFL1 & NPR1 & QMF1 \\                   
\hline
50       & 23   & 7    & 29   \\
\hline
55       & 31   & 7    & 41   \\
\hline
60       & 45   & 15   & 59   \\
\hline
65 		 & 65   & 23   & 83  \\
\hline
70 		 & 91   & 35   & 115 \\
\hline
\end{tabular}
\label{table_truncation_SIR}
\end{table}

The $70$ dB target SIR may be interesting to consider as it corresponds to the nearly perfect reconstruction case of the MMB4 PF.
However, this target requires a large number of coefficients ($35$ for NPR1). 
In practice, a SIR due to truncation of $55$ dB may be sufficient since channel impairments already degrade the resulting SIR, as illustrated in the next sub-sections.
For the rest of the paper, $N_G$ is chosen so that each PF has the same SIR of $55$ dB, enabling a fair comparison. 
Therefore, we have:

\begin{itemize}
\item $N_G = 31$ for TFL1.
\item $N_G = 7$  for NPR1.
\item $N_G = 41$ for QMF1.
\end{itemize}

The TFL1 and QMF1 PFs require more than $30$ non-zero coefficients to obtain this SIR target for a FS implementation. 
Such high number of coefficients may not be acceptable if a low complexity receiver is targeted. 
For the NPR1 PF, only $7$ coefficients are required to achieve a SIR up to $55$ dB, making it better suited for the FS implementation.  
It is worth noting that a convolution operation using $7$ coefficients may appear too complex to implement in practice. 
Therefore, a low-complexity hardware architecture is proposed in Section \ref{section_complexity_eval} to address this aspect.  

\subsection{Robustness to timing offset}

Timing offset impairment occurs when the transmitter and receiver baseband samples are not perfectly aligned in time. 
It is always the case in practice, since the channel introduces a propagation delay. 
Therefore, timing synchronization algorithms must be employed. 
In LTE uplink case, the timing synchronization is realized using time advance mechanism \cite{3GPP_36.133} to compensate the propagation delay of each User Equipment (UE) located at different geographical distance from the base station. 
However, new highly demanding scenarios like massive machine communications are considered in 5G. 
To reduce energy consumption and to improve spectral usage, time advance mechanism should be avoided and relaxed synchronization should be supported, where the propagation delay of each UE is not compensated. 
Therefore, synchronization errors appear, which causes two types of impairments:

\begin{enumerate}
\item Linear phase rotation for each sub-carrier due to the additional delay. This effect can be totally compensated after channel estimation and equalization. Indeed, if $l_d$ is the time offset in number of samples, then the frequency domain compensation term is expressed as $C_{TO}(m) = e^{-i\frac{2 \, \pi \, m \, l_d}{M}}$.
\item Inter-Symbol and Inter-Carrier Interference (ISI and ICI) due to PF misalignment between the transmitter and the receiver.
\end{enumerate}

It is considered, in this paper, that the OFDM signal is synchronized ($l_d = 0$) at the middle of its cyclic prefix. 
If $-\frac{L_{CP}}{2} < l_d < \frac{L_{CP}}{2}$, where $L_{CP}$ is the length of the cyclic prefix, then orthogonality is perfectly restored, since a circular shift in time domain represents a linear phase rotation in frequency domain. 
In 4G/LTE, $\frac{L_{CP}}{M} = 7\%$ for OFDM. 
Thus, orthogonality is still guaranteed if $|\frac{l_d}{M}| < 3.5\%$, where $|.|$ represents the absolute value operator.

Due to the absence of CP in a FBMC system, timing offset will result in unavoidable performance degradation. 
However, depending on the use of PPN or FS implementation, results are different due to the application of different timing offset compensation techniques. 
For the FS implementation, the compensation step lies between the FFT and the FS filtering stage, whereas in the PPN case it is performed after the PPN and the FFT. 

\begin{figure}[!t]
\centering
\normalsize
\includegraphics[width=3.5in]{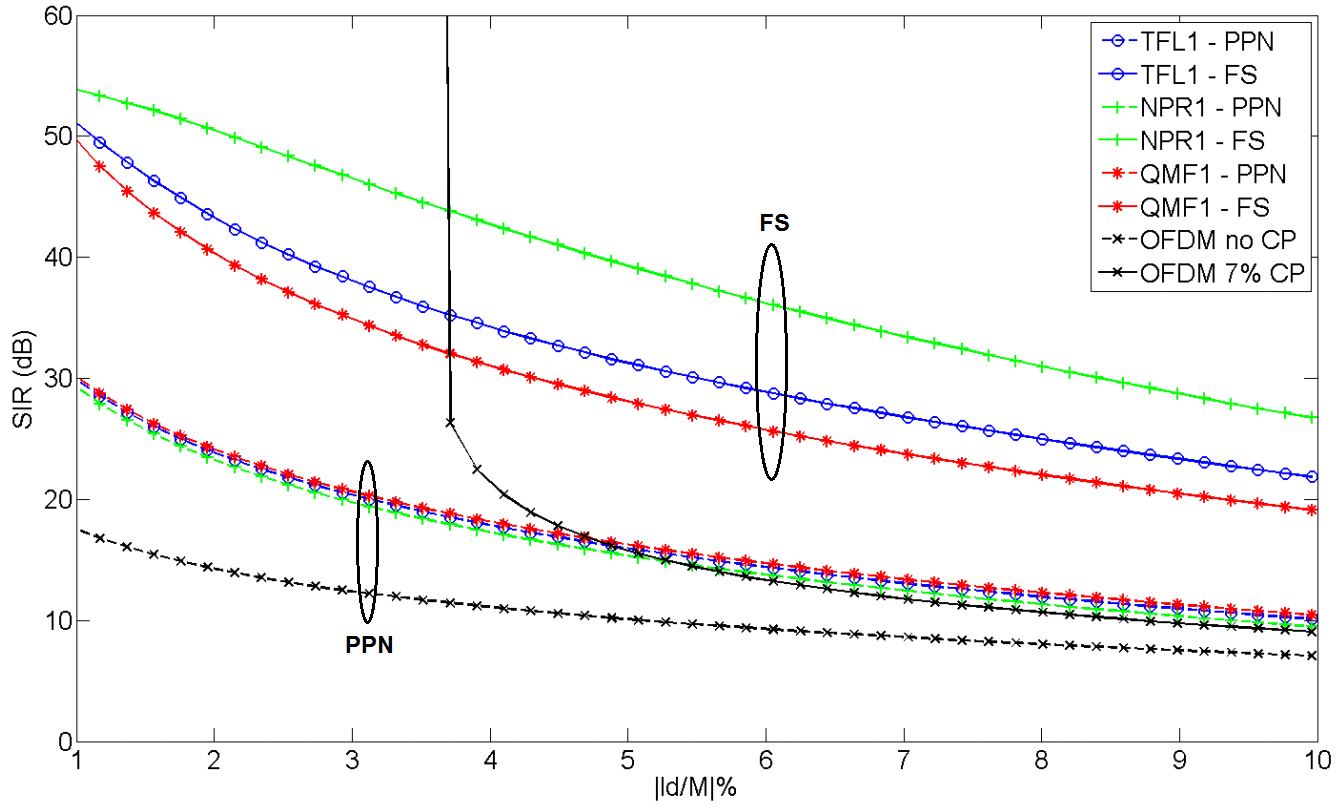}
\caption{Timing offset evaluation in terms of measured SIR for OFDM and FBMC with the considered short PFs. The effect of different implementations is also evaluated.}
\label{fig_TO_eval00}
\end{figure}

For the PPN-FBMC case, the SIR expression in (\ref{equation_recPAM2}) can be adapted to obtain the expression of the recovered PAM symbol when a timing offset of $l_d$ samples is applied, as follows:

\setlength{\arraycolsep}{0.0em}
\begin{eqnarray}
\hat{a}_n(m) && =  \Re \Biggl[ C_{TO}(m) \phi^*_n(m) \sum_{q=-Q}^{Q} \sum_{k=0}^{M-1} g(k) g \Bigl( k+q\frac{M}{2} + l_d \Bigr)\nonumber \\ 
             && \,\,\,\,\, \times \, x_{n-q}\Bigl(k+n\frac{M}{2} + l_d \Bigr) e^{-i2 \pi \frac{km}{M}} \Biggr]\nonumber \\ 
             && = \Re \Biggl[ \sum_{(p,q)\in \Omega} i^{- q - p} F_g(p,q\frac{M}{2}+l_d) \nonumber \\ 
             && \,\,\,\,\, \times \, a_{n-q}(m-p) e^{-i2\pi\frac{p l_d}{M}} \Biggr], \nonumber
\end{eqnarray}
\setlength{\arraycolsep}{5pt}

\noindent and the expression of the SIR becomes:
\setlength{\arraycolsep}{0.0em}
\begin{eqnarray}
\text{SIR}_{\text{PPN}}(l_d) && = \frac{ \Re \Bigl( F_g(0,l_d) \Bigr)^2 } { \sum_{(p,q) \in \Omega^0} \Re \Bigl[ i^{p+q} F_g(p,q\frac{M}{2}+ld) e^{-i2\pi\frac{p l_d}{M}} \Bigr]^2 }
\label{equation_SIR_PPN_TO}
\end{eqnarray}
\setlength{\arraycolsep}{5pt}
Note that the number of FBMC symbols acting as interference denoted by $Q$ must be set to $2$. 
Indeed, when a timing offset is considered, the FBMC symbols at $q = - M$ and $q = M$ are now acting as interference. 
Therefore, the obtained numerical values are similar to that obtained by simulation in Figure \ref{fig_TO_eval00}. 
Concerning the FS-FBMC receiver, it has been evaluated in \cite{MTB2015a}, where the following expression is obtained:
\setlength{\arraycolsep}{0.0em}
\begin{eqnarray}
\text{SIR}_{\text{FS}}(l_d) && = \frac{ 1 }{ \Bigl( \frac{\sigma_{INT}}{\sigma_{a}} \Bigr)^2 + 2\sum_{l=0}^{l_d-1} g^2(l) }
\label{equation_SIR_FS_TO}
\end{eqnarray}
\setlength{\arraycolsep}{5pt}
where $\sigma_{INT}^2$ is the power of the residual interference when $l_d = 0$.
In our case, the residual interference comes from the NPR nature of the PF and the truncation applied on the filter coefficients. 
Therefore, we have:
\setlength{\arraycolsep}{0.0em}
\begin{eqnarray}
\Bigl( \frac{\sigma_{INT}}{\sigma_{a}} \Bigr)^2 && = \sum_{(p,q)\in \Omega^0} \Re \Bigl[ i^{p+q} F_{g,g_t}(p,q) \Bigr]^2 \nonumber
\end{eqnarray}
\setlength{\arraycolsep}{5pt}
Figure \ref{fig_TO_eval00} shows SIR values for each considered short PF using the FS-FBMC receiver. 
The $N_G$ parameters used for the PFs are those defined in Sub-section \ref{subsec_truncated_impact}.
It is clear that independently from the used PF, the FS implementation outperforms the PPN implementation, a result that was already demonstrated for the case of the MMB4 PF \cite{BDN2014} and the QMF1 PF \cite{BMT2015}.
When using the PPN implementation, the timing offset error compensation is done in frequency domain, thus after the filtering stage. 
This lowers the compensation efficiency, causing ICI and ISI as mentioned above. 
In case of FS implementation, the compensation is more efficient since the filtering stage is performed in frequency domain, after compensation of the timing offset error. 
This explains the gap in performance between PPN and FS implementations. 

A gain of more than $10$ dB can be observed with NPR1 PF when compared to QMF1 PF for timing offset inferior to 5\%, and $8$ dB when compared to TFL1 PF. 
Around $3$ dB of difference is visible between TFL1 and QMF1 PFs. 
From (\ref{equation_SIR_FS_TO}), it is clear that the NPR1 achieves a higher SIR than the other PFs.
Figure \ref{figure_IR_FS_shape}b shows the impulse response of each PF. 
It can be observed that the amplitude of the NPR1 impulse response is lower at its edges. 
Therefore, the term $ \sum_{l=0}^{l_d-1} g^2(l) $ has the lowest value for NPR1, confirming its higher robustness against timing offset error than the other PFs.  

Concerning OFDM, it is clearly outperformed by FS-FBMC implementation when $|\frac{l_d}{M}| > 3.5\%$.
A gap of at least $20$ dB can be observed between FBMC with NPR1 PF and OFDM. For lower timing offset impairments, FBMC still exhibits acceptable performance since the SIR remains superior to $40$ dB for the NPR1 PF.

These results, validated by simulation, point out that the proposed NPR1 is the most interesting PF to combat timing offset impairment due to imperfect timing synchronization.
This is particularly interesting to fulfill the relaxed synchronization requirement foreseen in specific 5G communication scenarios like massive machine communications.

\subsection{Robustness to frequency offset}

\begin{figure}[!t]
\centering
\normalsize
\includegraphics[width=3.5in]{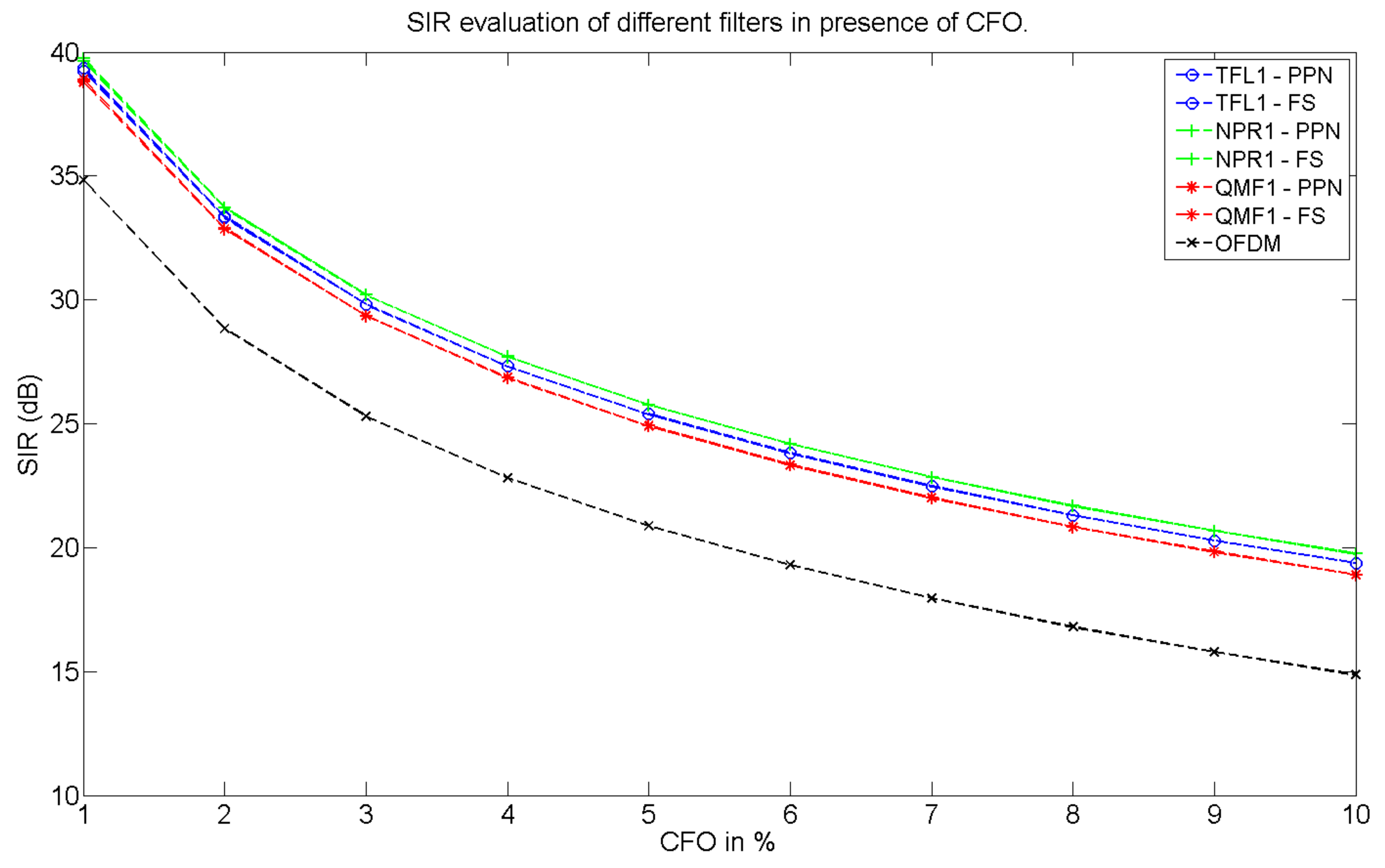}
\caption{SIR evaluation in presence of CFO.}
\label{fig_CFO_eval}
\end{figure}

Frequency offset impairment is a common issue in communication systems, and it is the consequence of the transmitter and/or receiver being in a situation of mobility (Doppler Shift/Spread). 
It also appears when there is a frequency misalignment in local oscillators of the transmitter and the receiver. 
Mathematically, it corresponds to a linear phase rotation of the received baseband samples. 
In 4G/LTE downlink case, the CFO is estimated and compensated in time domain by multiplying the received baseband samples by $e^{-i\frac{2 \, \pi \, k \, r}{M}}$, where $r \in \, ]-\frac{1}{2},\frac{1}{2}]$ is the CFO value relative to the sub-carrier spacing expressed as a percentage. 
However, in 4G/LTE uplink (and related 5G scenarios), it is not possible to compensate it directly in time domain since all baseband signals of all users overlap. 
In fact, it generates two types of impairments after demodulation:

\begin{enumerate}
\item Common Phase Error impairment (CPE). All the sub-carriers in a given symbol experience a phase rotation. The rotation angle is incremented at each received symbol. It can be easily compensated in frequency domain if the CFO is estimated, as the CPE term to compensate is  $C_{CPE}(n) = e^{i \pi n r}$. 
\item ICI due to misalignment of the transmitter and receiver PFs in frequency domain, also resulting in inter-user interference (IUI) in related 5G scenarios.
\end{enumerate}

The second described impairment represents a major issue, particularly for OFDM due to its low frequency localization. 
FBMC is naturally more robust against this type of ICI, especially when using a short PF \cite{LGS2014}. 
Therefore, it is expected that FBMC has higher robustness against CFO than OFDM. 
This is confirmed in Figure \ref{fig_CFO_eval}, which shows the SIR performance in presence of CFO with all the considered PFs, obtained both by numerical and simulation results. 
The SIR expression can be obtained by adapting equation (\ref{equation_recPAM2}), as follows: 
\setlength{\arraycolsep}{0.0em}
\begin{eqnarray}
\hat{a}_n(m) && =  \Re \Biggl[ C_{CPE}(n) e^{i\pi n r} \phi^*_n(m) \sum_{q=-Q}^{Q} \sum_{k=0}^{M-1} g(k) g \Bigl( k+q\frac{M}{2} \Bigr)\nonumber \\ 
             && \,\,\,\,\, \times \, x_{n-q}\Bigl(k+n\frac{M}{2}+ l_d \Bigr) e^{-i2 \pi \frac{k(m+r)}{M}} \Biggr]\nonumber \\ 
             && = \Re \Biggl[ \sum_{(p,q)} i^{- q - p} F_g(p+r,q\frac{M}{2})  \nonumber \\
             && \,\,\,\,\, \times \, a_{n-q}(m-p)  \Biggr]. \nonumber
\end{eqnarray}
\setlength{\arraycolsep}{5pt}
Assuming that the interference introduced by the truncation is negligible, the expression of the SIR for both FS-FBMC and PPN-FBMC receivers is:
\setlength{\arraycolsep}{0.0em}
\begin{eqnarray}
\text{SIR}_{\text{CFO}}(r) && = \frac{ \Re \Bigl( F_g(r,0) \Bigr)^2 } { \sum_{(p,q) \in \Omega^0} \Re \Bigl[ i^{p+q} F_g(r,q\frac{M}{2}) \Bigr]^2 }. \nonumber
\end{eqnarray}
\setlength{\arraycolsep}{5pt}
Up to $5$ dB of SIR can be observed between OFDM and FBMC with the NPR1 PF. 
This later is also the PF having the highest robustness against CFO. 
When compared to the TFL1 PF, a difference of $\approx 0.4$ dB is observed, and almost $0.9$ dB when compared to the QMF1 PF. 

For this particular channel impairment, PPN-FBMC and FS-FBMC receivers have similar performance. 
This can be explained by the fact that the compensation term $C_{CPE}$ only depends on the FBMC symbol index. 
Therefore, it can be integrated before and after the filtering stage without any mathematical difference. 
The only difference comes from the interference introduced by the filter truncation in FS-FBMC, however the impact on the SIR is negligible.

\subsection{Performance comparison over multipath channels}

\begin{figure}[!t]
\centering
\normalsize
\includegraphics[width=3in]{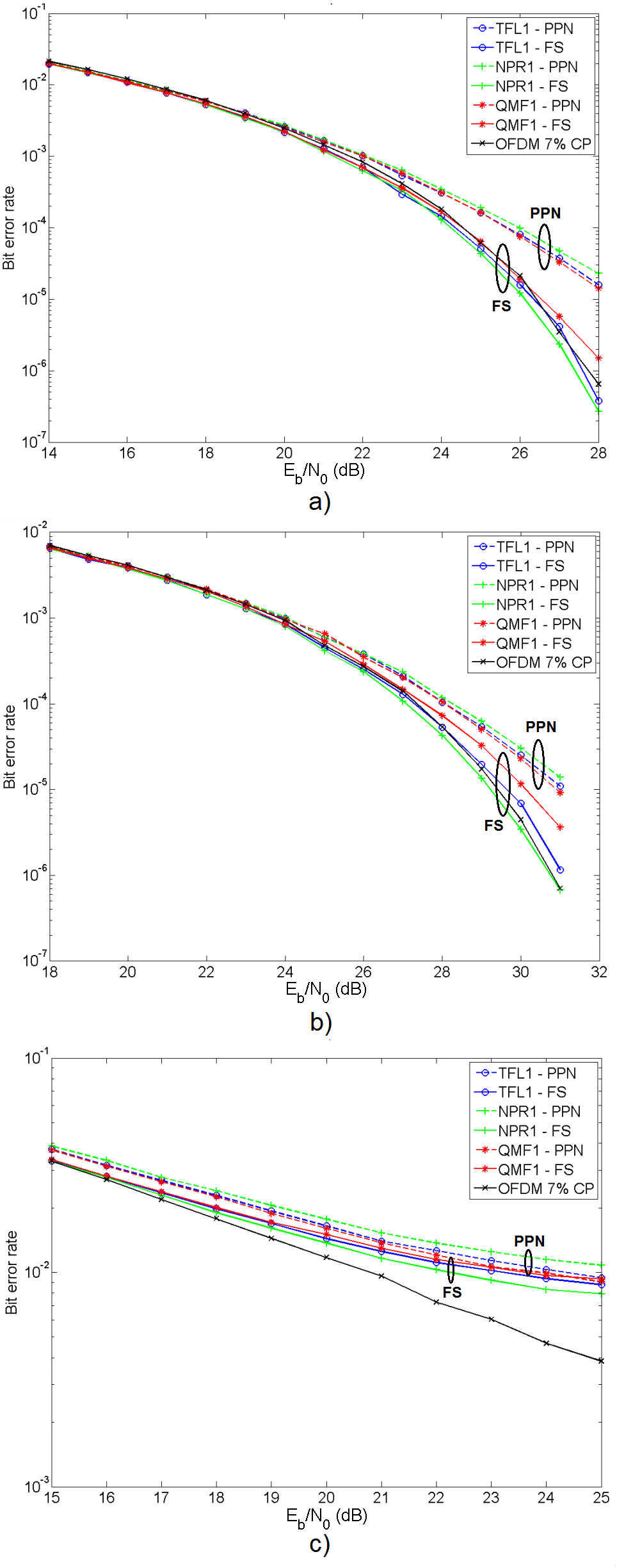}
\caption{BER evaluation of OFDM and FBMC with different PFs and implementations in presence of AWGN, for a) EPA static channel, b) EVA static channel, c) ETU static channel.}
\label{fig_EPA_EVA_ETU_eval}
\end{figure}

In the context of the 4G/LTE standard, three multipath fading channel models are defined \cite{3GPP_36.104}: 

\begin{itemize}
\item Extended Pedestrian A (EPA) model: $\tau = 410$ ns,
\item Extended Vehicular A (EVA) model: $\tau = 2510$ ns,
\item Extended Typical Urban (ETU) model: $\tau = 5000$ ns,
\end{itemize}

\noindent where $\tau$ corresponds to the delay spread of the multipath channel. The delay and power profiles of each channel model are detailed in  \cite{3GPP_36.104}. 
In the 4G/LTE standard, an OFDM symbol duration is always equal to $66.7\,\micro\second$  without CP, and the sub-carrier spacing is always equal to $15$ kHz.

This sub-section aims at evaluating the effect of these channels on the error rate performance of uncoded FBMC using different short PFs, with PPN and FS implementations. 
LTE parameters are considered for an IFFT length of $512$ and a 16-QAM constellation. Thus, $L_{CP} = 36$ for OFDM, and $300$ actives sub-carriers are used, corresponding to $25$ RBs. 
However, the frame structure of LTE is not perfectly respected, since Demodulation Reference Signals (DM RS) \cite{3GPP_36.211} are not transmitted, and the Channel State Information (CSI) is considered to be perfectly known. 
Note that the CSI needs to be estimated in practice by sending, for instance, coded auxiliary pilots \cite{CQJ+2016}.

For a fair comparison, the same equalization technique is used for OFDM and FBMC. 
The equalization step is realized after the computation of the FFT, in frequency domain. 
The output samples of the FFT are simply divided by the frequency response of the channel, realizing the classical low-complexity and per-sub-carrier Zero Forcing (ZF) equalizer. 

Static (no Doppler shift/spread) multipath channels with Additive White Gaussian Noise (AWGN) are considered to only evaluate the multipath and fading effect on the performance of OFDM and FBMC demodulators in terms of Bit Error Rate (BER). 
Figure \ref{fig_EPA_EVA_ETU_eval}a shows the BER performance when using EPA channel models, considering PPN and FS based FBMC with different PFs. 
As expected, the FS implementation outperforms the PPN implementation for all the considered PFs, particularly at higher SNR values.
A difference of at least one decade of BER can be observed at $E_b/N_0 = 28$ dB. 
Furthermore, the FS implementation with TFL1 and QMF1 PFs shows comparable performance to OFDM with CP. 
FS implementation with NPR1 offers slightly better results than OFDM at moderate Eb/No values ($E_b/N_0 >20$ dB), due to the absence of CP and its robustness against all the different types of timing impairments. 

Similar conclusions can be made for a channel with a longer delay spread like EVA, as shown in Figure \ref{fig_EPA_EVA_ETU_eval}b. 
However, an exception should be made for the QMF1 PF, since it exhibits a performance level inferior to OFDM. 
On the other hand, with the NPR1 PF, FBMC remains superior to OFDM even for an EVA channel. 

Due to the absence of a CP, FBMC seems to be more sensitive to long delay spread channels as it is the case for the static ETU channel model. 
Indeed, OFDM with CP outperforms FBMC on this type of channels when $E_b/N_0 > 17$ dB, as shown in Figure  \ref{fig_EPA_EVA_ETU_eval}c. 
At low $E_b/N_0$ values, the FS implementation is close to OFDM, and offers better results than the PPN implementation. 
NPR1 is again the most interesting short PF when using a FS implementation.

\begin{figure}[!t]
\centering
\normalsize
\includegraphics[width=3.5in]{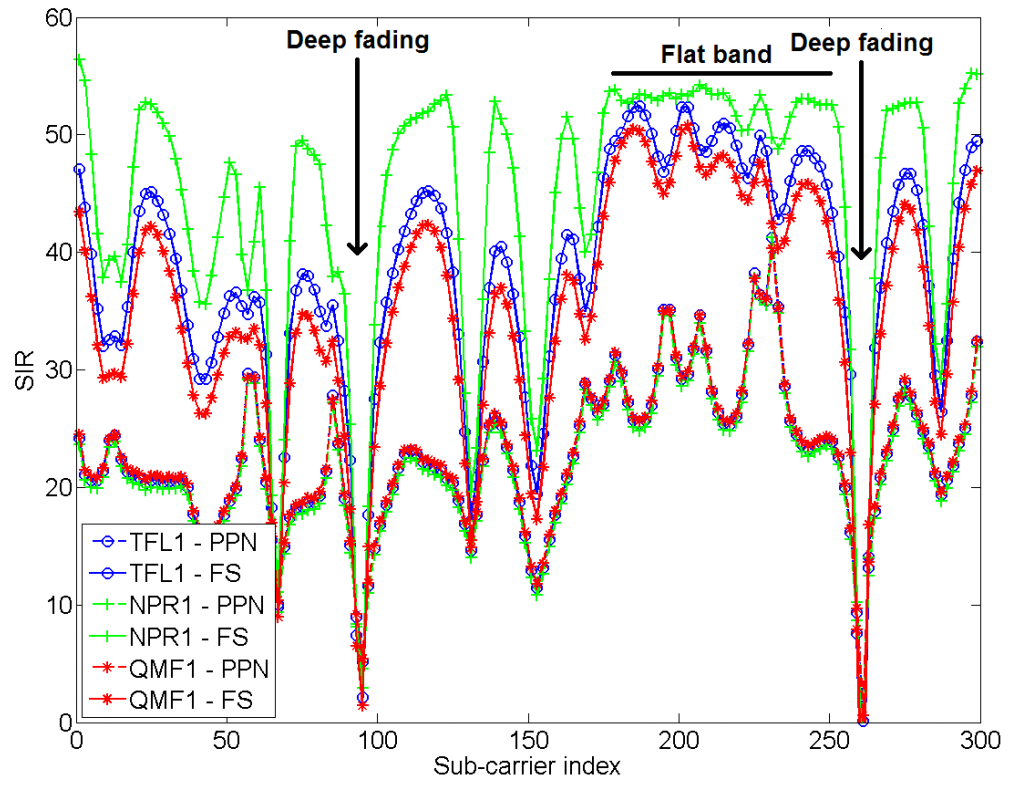}
\caption{SIR as function to the sub-carrier index for FS-FBMC and PPN-FBMC with different short PFs, for an ETU channel.}
\label{fig_SIR_ETU_eval}
\end{figure}

In fact, when deep fading occurs, the received signal is highly degraded, as shown in Figure \ref{fig_SIR_ETU_eval}.
This figure represents the SIR per sub-carrier with a randomly generated ETU channel, for different PFs and implementations. In the case of a flat fading in band, almost no interference occurs in FS implementation case (SIR $> 40$ dB). 
It is however not the case for the PPN implementation, where a gap of at most $30$ dB can be observed when compared to FS implementation, confirming the superiority of the FS implementation. 
Finally, as the delay spread of the ETU channel model being approximately two times longer than the delay spread of the EVA channel model, one straightforward solution is to double the duration of the FBMC symbol.

\section{Complexity evaluation}
\label{section_complexity_eval}

The FS-FBMC receiver is known to be more complex than the PPN-FBMC receiver. 
This is mainly due to the required convolution operation with $N_G$ truncated coefficients, compared to the simple windowing operation of the PPN-based implementation when short PFs are used. 
However, additional complexity reduction is possible for the FS-based receiver thanks to the properties of the PF and the OQAM scheme. 
After detailing the proposed complexity reduction approach, a hardware architecture is proposed for the FS filter stage and its complexity is evaluated and compared to the PPN filter stage.

\begin{figure*}[!t]
\centering
\normalsize
\includegraphics[width=7in]{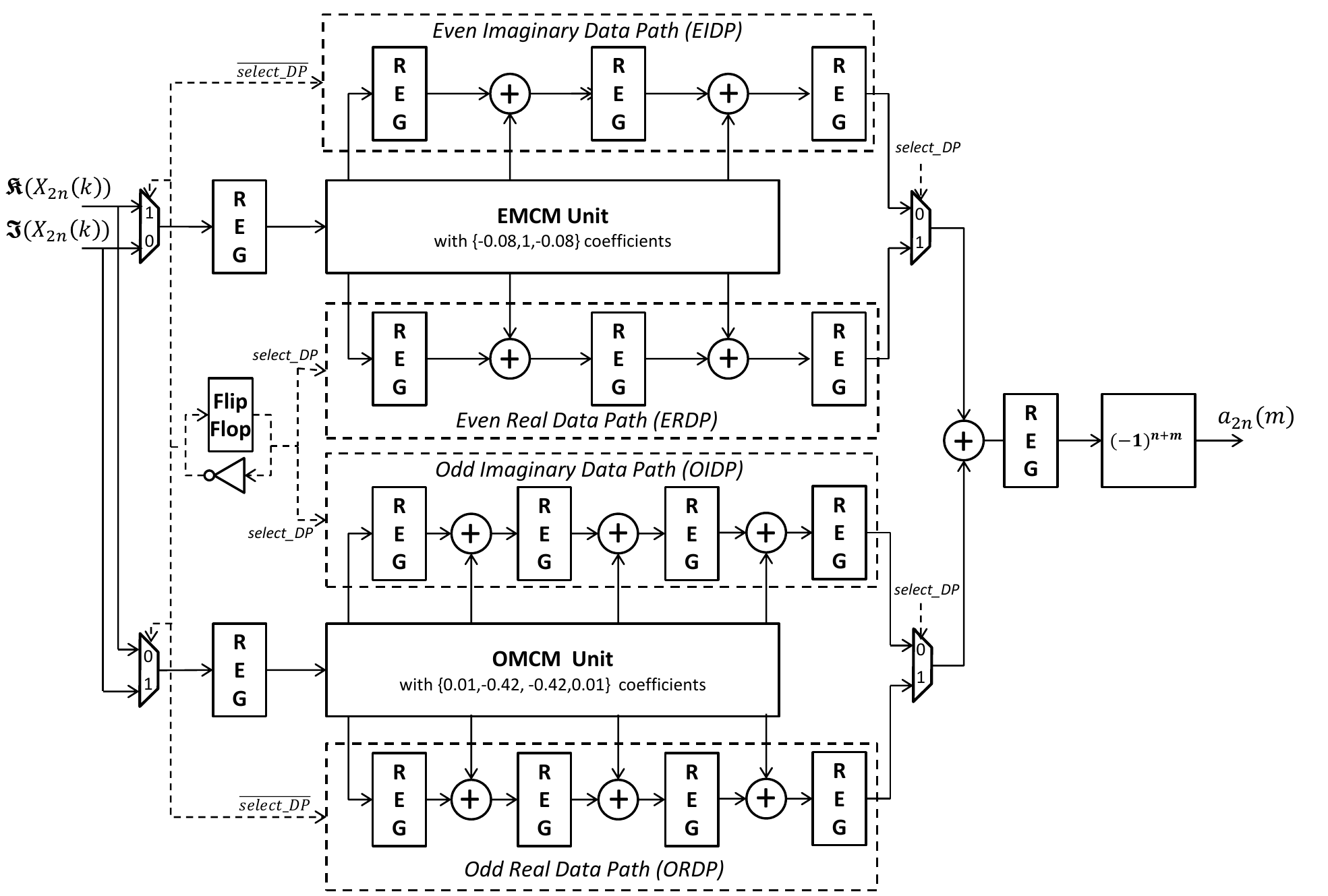}
\caption{Proposed FS filter stage architecture for NPR1 filter.}
\label{fig_filter_archi}
\end{figure*}

\subsection{Complexity reduction of the filter stage}
\label{subsec_complexity_reduction}
When a short PF is considered ($K=1$), the FFT size becomes equal to $M$, and there is no down-sampling step after the filter stage. 
Therefore, the complexity overhead comes from the circular convolution operation, which depends on the number of filter coefficients $N_G$ after truncation. 
A circular convolution typically requires $N_G$ Complex Multiplications (CMs) and $2(N_G-1)$ Real Additions (RAs) per sample.
However, these resources can be reduced by exploiting the properties of the PF and the OQAM scheme. 
First, if $g(k) = g(N-k)$ and $\Im(g)=0$, verified for the NPR1 PF (\ref{equation_g_NPR1}), then we have:
\begin{align}
 G(l) &= \sum_{k = 0}^{N/2-1} g(k) e^{i\frac{2\pi}{N}kl} + g(N-k) e^{i\frac{2\pi}{N}(N-k)l}  \nonumber \\
      &= \sum_{k = 0}^{N/2-1} 2g(k) \Re(e^{i\frac{2\pi}{N}kl}).
\end{align}
This expression shows that $G$ is real valued. 
Therefore, the filter stage requires now only $2N_G$ Real Multipliers (RMs) per sample. 
Furthermore, $G(-l)$ can be expressed as follows:
\begin{align}
 G(-l) &= \Bigl( \sum_{k = 0}^{N-1} g(k) e^{-i\frac{2\pi}{N}kl} \Bigr)^* \nonumber \\
       &=  G(l) \nonumber. \\
\end{align}
Consequently, the output of the filter stage $Y_{n}(m)$ becomes:
\begin{align}
\label{eq_secV_01}
Y_{n}(m) & = G(0) X_{n}(m) + \sum_{l=1}^{\Delta+1} G(l) \Bigl( X_{n}(m-l) + X_{n}(m+l) \Bigr) \nonumber \\
		 &= \sum_{l=0}^{\Delta+1} G(l) X_{n}(m,l),  \nonumber  \\
\end{align}
where $X_{n}(m,l) = X_{n}(m-l) + X_{n}(m+l)$ if $l>0$ and $X_{n}(m,0) = X_{n}(m)$. 
One RM can be removed by re-scaling the PF coefficients by $G'(l) = G(l)/G(0)$. Then, (\ref{eq_secV_01}) becomes:
\begin{align}
\label{eq_secV_02}
Y_{n}(m) = \sum_{l=0}^{\Delta+1} G'(l)  X'_{n}(m,l).   \nonumber \\
\end{align}
with $X'_{n}(m,l) = G(0) X_{n}(m,l)$. 
This scaling factor $G(0)$ can be integrated without complexity increase in the equalizer coefficients. 
Alternatively, it can be taken into account in the decision stage (QAM demapper). 
The re-scaled PF coefficients $G'(l)$ can be computed during design time, and stored in a Look-Up-Table (LUT). 
At this step, the filter stage requires $2\Delta$ RMs per sample. 
This number can be further reduced since half of the circular convolution outputs are discarded due to the OQAM scheme.
In this case, we have:

\begin{align}
\hat{a}'_{2n}(2m)     &= (-1)^{(n+m)}   \sum_{l=0}^{\Delta+1} G'(l) \Re \Bigl( X'_{2n}(2m,l) \Bigr),   \nonumber \\
\hat{a}'_{2n}(2m+1)   &= (-1)^{(n+m+1)} \sum_{l=0}^{\Delta+1} G'(l) \Im \Bigl( X'_{2n}(2m+1,l) \Bigr),   \nonumber \\
\hat{a}'_{2n+1}(2m)   &= (-1)^{(n+m+1)} \sum_{l=0}^{\Delta+1} G'(l) \Im \Bigl( X'_{2n+1}(2m,l) \Bigr),   \nonumber \\
\hat{a}'_{2n+1}(2m+1) &= (-1)^{(n+m+1)} \sum_{l=0}^{\Delta+1} G'(l) \Re \Bigl( X'_{2n+1}(2m,l) \Bigr), 
\label{eq_secV_03} 
\end{align}

\noindent where $\hat{a}'_n(m)$ being the re-scaled $\hat{a}_n(m)$ signal. 
Therefore, there is no need to process $\Im \Bigl( X'_n(m,l) \Bigr)$ or $\Re \Bigl( X'_{n}(m,l) \Bigr)$ depending on the parity of $n$ and $m$. 
Only $\Delta$ RMs and $2\Delta$ RAs per sample are now required. 
Additionally, only the outputs corresponding to the allocated sub-carrier indexes can be considered.
If the number of allocated sub-carriers is denoted by $N_c$, this gives $\Delta N_c$ RMs required per FBMC symbol.

Regarding the PPN stage, it requires $2M$ RMs per FBMC symbol for short PFs. 
Thus, when considering only multiplication operations, the complexity ratio between FS and PPN filter stages is given by $R_{RM} = \alpha \Delta / 2$, with $\alpha = N_c/M \approx 0.586$ in 4G/LTE. 
For QMF1 ($\Delta = 20$) and TFL1 ($\Delta = 15$) PFs, the complexity (in number of RMs) is multiplied by $5.86$ and $4.395$ respectively. 
This confirms that these PFs are not suitable for the FS implementation. 
However, for NPR1 PF ($\Delta = 3$), the FS filter stage is $12\%$ less complex than the PPN stage. 

It is worth noting that the PPN stage requires a LUT memory of depth $M$ to store the PF coefficients. 
Furthermore, the FS filter stage still require $2 \Delta N_c$ additions per FBMC symbols. 
When targeting hardware implementation, registers must also be considered to store the input signal $X'_n(k)$ due to the iterative processing of the circulation convolution operation.
Finally, this operation uses constant filter coefficients which do not change during the processing iterations. This particularity can be taken into account to further reduce the complexity when considering hardware implementation.

Therefore, the above ratio may not be accurate enough to reflect the comparative hardware complexity since it only considers the number of multipliers. 
For an accurate comparison, we propose a detailed hardware architecture for the FS filter stage.

\subsection{Proposed hardware architecture for the FS filter stage}
\label{subsec_hardware_archi}

The circular convolution operation can be efficiently implemented in hardware using a typical FIR filter architecture. 
Such architecture can take one input and generate one output per clock cycle in pipelined manner.
If a Multiple Constant Multiplier (MCM) is used, multiplier-less FIR architecture can be designed for fixed-point precision. 
If $G_Q$ are the PF coefficients quantized on $Q$-bit to be multiplied by a $Q$-bit input $X_Q$, then:

\begin{align}
G_Q \times X_Q = \sum_{k=0}^{Q-1} c_k 2^k X_Q,   \nonumber \\
\end{align}

\noindent where $c_k \in \{-1,0,1\}$ denotes the symbol number $k$ of the Canonical Signed Digit (CSD) representation of $G_Q$ \cite{HS2000}. 
Therefore, only adders and registers are required for this architecture. 
It is advantageous to consider it if the same set of coefficients has to be re-used for each processed sample, which is the case for the FS filter stage. 
As an adder requires less hardware resources (logic gates) than a multiplier, important hardware complexity reduction is expected.

As a baseline solution, (\ref{eq_secV_02}) can be directly implemented using one FIR filter for the computation of the real part, and another FIR filter for the computation of the imaginary part. 
It is however not an optimal choice if a low-complexity implementation is targeted, since half of the generated samples by both FIR filters are discarded due to the OQAM scheme, as shown in (\ref{eq_secV_03}). 
However, this equation cannot be directly implemented using a typical FIR filter. 
In fact, the content of the corresponding registers must be switched between the real and imaginary parts of the stored $X'_n(m)$ samples.
Therefore, we propose a novel architecture adapted for the FS filter stage.

In the following, only even FBMC symbol indexes are considered ($\hat{a}'_{2n}(m)$ and $X'_{2n}(m)$). Similar demonstration can be applied for the odd indexed symbols. Index $2m$ of (\ref{eq_secV_03}) can be rewritten as follows:

\begin{align}
\hat{a}'_{2n}(2m) = & (-1)^{(n+m)} \Bigl(  \sum_{l=0}^{\lfloor \frac{ \Delta+1 }{2} \rfloor } G_{even}(l) \Re \Bigl( X'_{2n}(2m,2l) \Bigr)   \nonumber \\
			 & + \sum_{l=0}^{ \lfloor \frac{\Delta+1}{2}  \rfloor } G_{odd}(l) \Re \Bigl( X'_{2n}(2m,2l+1) \Bigr)  \Bigr),
\end{align}

\noindent where $G_{even}(l) = G(2l)$ and $G_{odd}(l) = G(2l+1)$. Similarly, index $2m+1$ of (\ref{eq_secV_03}) becomes:

\begin{align}
\hat{a}'_{2n}(2m+1) = & (-1)^{(n+m+1)} \Bigl( \nonumber \\
                & \sum_{l=0}^{\lfloor \frac{ \Delta+1 }{2} \rfloor } G_{even}(l) \Im \Bigl( X'_{2n}(2m+1,2l) \Bigr)   \nonumber \\
		     & + \sum_{l=0}^{ \lfloor \frac{\Delta+1}{2}  \rfloor } G_{odd}(l) \Im \Bigl( X'_{2n}(2m+1,2l+1) \Bigr)  \Bigr).
\end{align}

The above equations show that the FS filter stage can be separated into two FIR filters, each respectively holding even and odd indexes of the PF coefficients. 
Indeed, for each received $X'_{2n}(2m)$ sample, its real part is processed by the even-indexed FIR filter, while its imaginary part is processed by the odd-indexed FIR filter. 
Conversely, the real part of $X'_{2n}(2m+1)$ is processed by the odd-indexed FIR filter, and its imaginary part by the even-indexed FIR filter. 
Therefore, $2$ FIR filters are required, similarly to the baseline solution. 
However, the number of required coefficients per FIR filter is divided by $2$, reducing the complexity. 
The obtained architecture using the NPR1 filter is presented in Figure~\ref{fig_filter_archi}.

The Even MCM (EMCM) unit generates the multiplications by the even-indexed filter coefficients, while the Odd MCM (OMCM) unit generates the multiplications by the odd-indexed filter coefficients.
The behaviour of the architecture executes in two phases, which are repeated continuously. 
Each phase takes one clock cycle.

In the first phase, the real part of $X'_{2n}(2m)$ is sent to the EMCM unit while its imaginary part is sent to the OMCM unit. 
Meanwhile:

\begin{itemize}
\item the registers of the Even Real Data Path (ERDP) (Figure~\ref{fig_filter_archi}), belonging to the even-indexed FIR filter, are enabled by the \emph{select\_DP} control signal,
\item the registers of the Odd Imaginary Data Path (OIDP), belonging to the odd-indexed FIR filter, are also enabled,
\item the registers of the Even Imaginary Data Path (EIDP) and the Odd Real Data Path (ORDP) are both disabled.
\end{itemize}

Finally, the outputs of the ERDP and the ORDP are summed together. 
Furthermore, sign inversion is performed depending on the $(-1)^{n+m}$ term value of (\ref{eq_secV_03}).

At the second phase, the real part of $X'_{2n}(2m)$ is sent to the OMCM unit while its imaginary part is sent to the EMCM unit. 
The registers which were disabled (respectively enabled) are now enabled (respectively disabled) by the \emph{$\overline{\text{select\_DP}}$} control signal. 
The outputs of the EIDP and the OIDP are selected and summed together, followed by a possible sign inversion depending on the $(-1)^{n+m+1}$ term value.

\subsection{Hardware complexity comparison}

The proposed FS filter stage architecture, in addition to the baseline architecture, were described in VHDL/Verilog and synthesized targeting the XC7z020 Xilinx Zynq SoC device. 
All the MCM units are generated using the SPIRAL code generator \cite{PMK2005+}. The results, summarized in Table \ref{table_hardware_comparison}, include:

\begin{itemize}
  \item the PPN unit detailed in \cite{NNB2015},
  \item the baseline FS filter directly derived from equation (\ref{eq_secV_02}),
  \item the proposed FS filter presented in the previous sub-section.
\end{itemize}

The architecture of the PPN unit in \cite{NNB2015} is adapted to process $2$ FBMC symbols in parallel (OQAM scheme). 
To enable a fair comparison, we have considered $2$ FS filter stages in parallel, in such a way that the same processing speed is achieved. 
Furthermore, the same quantization chosen in \cite{NNB2015} is considered:

\begin{itemize}
  \item the samples at the input and the output of each unit use 16-bit quantization,
  \item all filter coefficients are quantized on 12-bits.
\end{itemize}

Only the NPR1 PF is considered in this section. 
The QMF1 and TFL1 PFs are less adapted for FS implementation since they require, at least, $4$ times more filter coefficients (see Sub-section \ref{subsec_complexity_reduction}).

The baseline solution of the FS filter stage requires $11.5$\% less LUTs than the PPN unit. 
This confirms that one MCM is less complex than $2$ multipliers. 
On the other hand, the FS filter stage uses $2.85$ times more flip-flops due to the FIR filter registers. 
If we consider that LUTs and flip-flops have similar complexity, then the baseline FS filter stage requires $13$ \% more hardware resources than the PPN unit. 
It also achieves a clock frequency speed of $186$ MHz, $15$ \% less than the PPN unit. 

Concerning the proposed architecture, it requires $38$\% less LUTs than the PPN unit and $30$\% less LUTs than the baseline solution. 
The number of required flip-flops is almost unchanged when compared to the baseline solution. 
However, the proposed architecture is less complex than the PPN implementation since it requires $11$\% less in total hardware resources. 
It also achieves a clock frequency speed of $218$ MHz, which is $17$\% higher than the baseline solution.

Section \ref{Section_Performance_evaluation_of_short_filter} shows that the FS-FBMC receiver offers improved robustness when compared to PPN-FBMC against timing offset and multi-path impairments (assuming ZF equalizer is used). 
Furthermore, hardware complexity evaluation conducted in this section shows that the proposed FS filtering stage (using the proposed NPR1 PF) has a lower complexity than the PPN implementation. 
This concludes that the FS-FBMC receiver is more advantageous to use than the PPN-FBMC receiver  when using the proposed short NPR1 filter.

\begin{table}[!t]
\caption{Required hardware resources for each considered unit.}
\centering
\begin{tabular}{|c|c|c|c|c|}
\hline
\textbf{Filter stage}  &  \textbf{LUTs} & \textbf{Flip-Flops} &  \textbf{Total} & \textbf{Frequency} \\   
\hline
PPN unit \cite{NNB2015}         & $1460$  & $135$ & $1595$  & $220$ MHz \\
\hline
Baseline FS filter stage        & $1292$  & $512$ & $1804$  & $186$ MHz\\
\hline
Proposed FS filter stage        & $902$   & $520$ & $1422$  & $218$ MHz\\
\hline
\end{tabular}
\label{table_hardware_comparison}
\end{table}

\section{Conclusion}
\label{Section_Conclusion}

In this paper, a novel short PF (NPR1) suitable for several 5G scenarios is proposed. 
In presence of timing offset due to imperfect synchronization, the NPR1 PF, combined with the FS implementation, exhibits a gain of more than 8 dB of SIR when compared to SoTA short PFs (TFL1 and QMF1).
It outperforms OFDM, where a gap of $20$ dB of SIR can be observed. 
The NPR1 PF is also the most robust filter to combat CFO. In the case of 4G/LTE multipath channel, the NPR1 PF is even better than OFDM for the EPA channel model, due to the absence of CP. 
In the case of ETU channel model, the NPR1 PF shows improved performance when compared to other FBMC PFs. 
Finally, an efficient hardware architecture of the FS filter stage is proposed. 
Hardware complexity evaluation shows that the proposed FS-based FBMC receiver, using the NPR1 PF, requires $11$\% less hardware resources than the PPN-based FBMC receiver. 
Therefore, combining the proposed NPR1 filter and FS-FBMC receiver architecture provides an original solution that combines complexity reduction and performance improvement with respect to a typical PPN-FBMC receiver.

\ifCLASSOPTIONcaptionsoff
  \newpage
\fi


\bibliographystyle{IEEEtran}
\bibliography{bibliography}

\end{document}